\documentclass[aps,prb,preprint,showpacs,showkeys]{revtex4}
\usepackage{amsmath}
\usepackage{graphicx}
\usepackage{color}

\begin{document}


\title{A bead on a hoop rotating about a horizontal axis: a 1-D ponderomotive trap}


\author{A. K. Johnson}
\affiliation{University of Illinois, Department of Physics, Chicago, IL 60607}

\author{J. A. Rabchuk}
\email[contact author ]{ja-rabchuk@wiu.edu}
\homepage[]{http://www.wiu.edu/users/mfjar2}
\affiliation{Western Illinois University, Department of Physics, Macomb, IL 61455}


\date{\today}

\begin{abstract}
We describe a simple mechanical system that operates as a ponderomotive particle trap, consisting of a circular hoop and a frictionless bead, with the hoop rotating about a horizontal axis lying in the plane of the hoop. The bead in the frame of the hoop is thus exposed to an effective sinusoidally-varying gravitational field. This field's component along the hoop is a zero at the top and bottom. In the same frame, the bead experiences a time-independent centrifugal force that is zero at the top and bottom as well. The system is analyzed in the ideal case of small displacements from the minimum, and the motion of the particle is shown to satisfy the Mathieu equation. In the particular case that the axis of rotation is tangential to the hoop, the system is an exact analog for the rf Paul ion trap. Various complicating factors such as anharmonic terms, friction and noise are considered. A working model of the proposed system has been constructed, using a ball-bearing rolling in a tube along  the outside of a section of a bicycle rim. The apparatus demonstrates in detail the operation of an rf Paul trap by reproducing the dynamics of trapped atomic ions and illustrating the manner in which the electric potential varies with time.

\end{abstract}


\maketitle


\section{\label{sec:intro} INTRODUCTION}

In a ponderomotive particle trap, a time-varying and spatially-inhomogeneous field produces an average restoring force on the particle that pushes it toward the minimum in the field amplitude. Such traps have been developed using both radio-frequency (rf) electric and magnetic fields to great advantage in trapping atomic ions and subatomic charged particles. In particular, the growing usefulness of the rf Paul trap for applications in mass spectrometry \cite{cit:massspec} and quantum information storage and processing \cite{cit:quinfo} has generated significant interest in finding a simple, mechanical demonstration of its principles.

What makes the principle of such traps less than obvious is that at any given instant there is always one direction along which the trapping field is pushing the particle away from the field minimum. Alternatively, along any axis through the trap center the field will vary sinusoidally between pushing the particle towards the minimum and pushing it away. The net difference between those two pushes is the ponderomotive force.  At sufficiently high driving frequencies, the ponderomotive force and the time-dependent field force operate on such different time scales that they are decoupled and can be treated independently, in what is known as the pseudo-potential approximation.\cite{ref:dehmelt} The time-averaged ponderomotive force gives rise to a pseudo-potential that controls the particle's overall behavior. This dominant but slower motion is called the secular motion, while the smaller amplitude and more rapid motion driven by the field is called the micromotion. When the maximum field amplitude is directly proportional to the distance from the minimum, as in the case of the hyperbolic rf Paul trap, the motion of the trapped particle is described exactly by the solution to a Mathieu equation.\cite{ref:Paul}

Ponderomotive traps involving the earth's constant gravitational field require a moving constraint, as in the case of the vertically-driven (inverted) pendulum or a ball on a rotating saddle surface.\cite{ref:vpend,ref:saddle} Wolfgang Paul used a ball on a rotating saddle-shaped platform to illustrate the mechanism of the Paul trap in his Nobel prize acceptance speech.\cite{ref:Paul} This device was mentioned by Rau as an example of the more general type of stability that can be mapped onto asymmetric rotors.\cite{ref:Rau} As noted by several authors who have recreated Paul's demonstration, however,\cite{ref:saddle,ref:monroe} this device can be tricky to construct, and the behavior of the ball in the trap is sometimes hard to control, perhaps due to friction, rolling and the accelerating surface. It is also a complicated problem to develop a mathematical description of the motion of the rolling ball along the rotating surface. Landau and Lifshitz showed that a pendulum driven sinusoidally at its pivot in the vertical direction satisfies the Mathieu equation.\cite{ref:lanlif} The inverted pendulum was suggested specifically as a model that illustrates the behavior of ions in rf traps.\cite{ref:vpend}  However, because the ponderomotive force is a complicated vector sum of gravity and the constraint force acting along the pendulum rod, the inverted pendulum is not very satisfactory as a demonstration of the ponderomotive trapping mechanism. What is more, the stability characteristics of the inverted pendulum differ from that seen in rf Paul traps.\cite{ref:vpend}

In this paper, a new approach for creating a one-dimensional, gravitational ponderomotive trap is developed. The behavior of a frictionless bead on a circular hoop rotating about a vertical axis has been studied as an example of constrained rotational motion.\cite{ref:Taylor} If this system is rotated instead about a horizontal axis passing through the plane of the hoop, the top and bottom points of the hoop become possible centers for ponderomotive traps. This is so first of all because the tangential component of gravity along the hoop vanishes at the top and bottom of the hoop. Secondly, as the hoop rotates, the gravitational field component tangent to the hoop alternates between pointing toward and away from either equilibrium point on the hoop. A constant, but spatially inhomogeneous, centrifugal force-`field' is also acting on the bead in the frame of the rotating hoop. This centrifugal force is zero at the equilibrium points and becomes increasingly significant the farther the bead is away from the axis of rotation, and the faster the hoop is rotating.

In this paper, the general equation of motion for the bead in the rotating hoop is obtained and analyzed. Then the bead's motion along the rotating hoop is shown to satisfy Mathieu's equation for small displacements and moderate angular frequencies. In the special case that the rotational axis passes through an equilibrium point, the static centrifugal field is eliminated and the mechanical trap simulates the behavior along a single axis of a pure rf Paul trap. The complicating factors of higher angular frequencies, larger initial displacements from equilibrium, friction, noise, and rolling rather than sliding motion are also discussed. A mechanical demonstration model for this trap has been constructed and tested. This demonstration vividly illustrates the trapping mechanism at work in rf Paul traps and successfully reproduces the Mathieu-type motion expected from a true ponderomotive trap. Finally, several problems that can be assigned for upper level undergraduates based on this paper are formulated in an appendix.

\section{\label{sec:theory} THEORY}


The proposed system for study is depicted in
\begin{figure}
\includegraphics[width=2.5in]{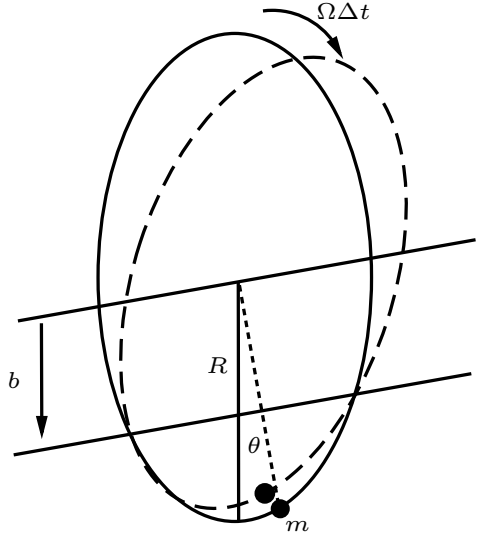} \label{fig:modeldiag} 
\caption{The hoop trap model. The rotating hoop has radius $R$, with a mass $m$ displaced from the equilibrium point on the hoop by an angle $\theta$. The dashed hoop is rotated from the upright position by an angle $\Omega \Delta t$ about a horizontal axis a distance $b$ below its diameter in the plane of the hoop.}
\end{figure}
Fig.~(\ref{fig:modeldiag}). The motion of the bead will be analyzed in the frame of the rotating hoop of radius $R$, and treated as a function of $\theta$, the angle of displacement from the equilibrium point on the hoop. The hoop is rotated about a horizontal axis in the plane of the hoop a distance $b$ below its diameter, such that the equilibrium point starts out at time $t = 0$ at the lowest point of the hoop.  The bead of mass $m$ is constrained to move along the hoop. Its distance from the axis of rotation is $R\cos\theta -b.$ The  kinetic energy and potential energy of the bead will be defined as functions of the variables $\theta$ and $\frac{d\theta}{d t} = \dot{\theta}.$

The kinetic energy of the bead has two terms, that of the motion of the hoop itself at the location of the bead and the motion of the bead along the hoop. Thus,
\begin{equation}\label{eq:kinen}
    T = {\scriptstyle \frac{1}{2}}  m (R \cos \theta - b)^{2} \Omega^{2} + {\scriptstyle \frac{1}{2}}  m R^{2} \dot{\theta}^{2}
\end{equation}
The potential energy $U$ of the bead relative to its energy at the level of the axis of rotation is given by
\begin{equation}\label{eq:poten}
    U = -m g (R \cos \theta - b) \cos \Omega t,
\end{equation}
where $g$ is the gravitational acceleration at the earth's surface. Therefore, the Lagrangian ${\cal L} = T - U $ becomes
\begin{eqnarray}\label{eq:Lagr}
    {\cal L} &=& {\scriptstyle \frac{1}{2}} m (R \cos \theta - b)^{2} \Omega^{2} + {\scriptstyle \frac{1}{2}} m R^{2} \dot{\theta}^{2} \\
&+& m g (R \cos \theta - b) \cos \Omega t.\nonumber
\end{eqnarray}
The equation of motion for the bead is obtained by requiring the Lagrangian function in Eq. (\ref{eq:Lagr}) to satisfy the Euler-Lagrange equation, $\partial{\cal L}/\partial \theta = (d/dt) (\partial{\cal L}/\partial \dot{\theta}).$ This results in a second-order differential equation for $\theta$,
\begin{equation}\label{eq:Fullsln}
    \ddot{\theta} = -\left[\Omega^{2} \left((R \cos \theta - b)/R \right) + (g/R) \cos \Omega t\right] \sin \theta.
\end{equation}

The two causes of the acceleration of the bead along the hoop arise from the gravitational force and from the centrifugal force due to the rotation of the hoop. The components of both of these forces acting along the hoop are proportional to ($\sin \theta$), and are zero at the top and bottom of the hoop. Therefore, the top and bottom of the hoop are equilibrium points. Since the magnitude of the centrifugal force depends on the distance from the bead to the axis of rotation, the centrifugal force component has an additional ($\cos \theta$) dependence. The centrifugal force always points away from the axis of rotation and toward the equilibrium point as long as the axis of rotation passes along a chord of the circular hoop. If the axis of rotation does not pass through the hoop, the centrifugal force points away from the nearest equilibrium point and toward the farthest one.  The gravitational force (in the frame of the hoop) is seen to alternate between pointing toward and away from each of the equilibrium points. The constant $g/R$ is the square of the natural frequency of the bead's oscillation about the lowest equilibrium point if the hoop were to be held at rest ($\Omega=0$), that is,
\begin{equation}\label{eq:natfreq}
g/R = \omega_0^2.
\end{equation}

\subsection{\label{sec:smallanglegen} The small angle limit}

For stable motion, oscillations of the bead along the hoop will be minimal and consequently the small angle approximation can be made, where $\cos \theta \approx 1$ and $\sin \theta \approx \theta$.  Making these substitutions, Eq.(\ref{eq:Fullsln}) becomes
\begin{equation}\label{eq:Smang}
   \ddot{\theta} \approx -\left[\Omega^{2} \left((R - b)/R\right) + \omega_0^2 \cos \Omega t\right] \theta.
\end{equation}
This equation of motion has the character of Hooke's Law, except that the `linear' restoring force has a sinusoidal time dependence. In fact, this equation can be transformed into a Mathieu equation.

\begin{figure}
 \includegraphics[width=2.5in]{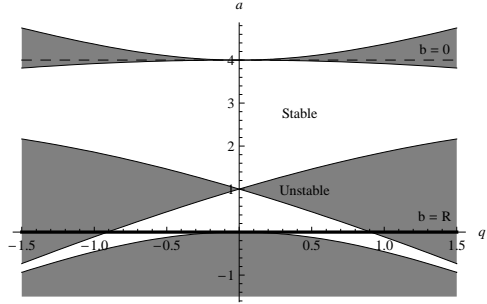}
 \caption{Stability plot of the Mathieu parameters $a$ vs. $q$. Stable regions of motion are unshaded and unstable regions of motion are shaded. The symmetry of the stability curves about the $a$-axis means that the sign of $q$ is unimportant. The portion of this parameter space explored by the bead on the rotating hoop depends on the value of $b$, the distance the horizontal axis of rotation is below the hoop center, and $\Omega$, the angular frequency of rotation. For a fixed value of $b$ the possible values for the parameters fall on a horizontal line in $a$ - $q$ space. As the frequency $\Omega$ is increased, the value of $q$ decreases until the limit of $q=0$ is reached for an infinitely large $\Omega.$ The dashed line (- - -) indicates the possible values of $a$ and $q$ when the axis of rotation passes through the middle of the hoop, and the solid line (---) indicates the possible values of $a$ and $q$ when the axis is tangent to the bottom of the hoop.} \label{fig:mathieuPlot}
\end{figure}

The canonical form of the Mathieu equation is defined as\cite{ref:Ruby}
\begin{equation}\label{eq:Mathieueqn}
   \frac{d^{2}Y}{dX^{2}} = -[a - 2q \cos 2X]Y
\end{equation}
with $a$ and $q$ representing real constants. Broadly speaking, the solutions to the Mathieu equation can separated into two types, depending on the values of the parameters $a$ and $q$.\cite{ref:McLachlan1} One type are the bounded, stable solutions for which the displacements from equilibrium always remain small. The other type are those solutions for which the oscillations of the system grow in amplitude until the displacements become infinitely large. These two classes of solutions are divided from each other in $a$ - $q$ parameter space by those curves that mark the $\pi$- or $2\pi$-periodic solutions. A plot indicating the regions of stability for the Mathieu equation is given in Fig.~(\ref{fig:mathieuPlot}). Those regions in $a$ - $q$ space for which the solutions to the Mathieu equation are unstable are colored gray, while the white regions indicate those values of the parameters for which the solutions are stable. The stability plot is symmetric about the $a$-axis, making the sign of the $q$-parameter irrelevant.

The equation of motion for the bead in the small angle approximation, Eq.~(\ref{eq:Smang}), can be rewritten in the canonical form of the Mathieu equation by making the substitution $2X=\Omega t$
\begin{equation}\label{eq:SmangMathieu}
   \frac{d^{2}\theta}{dX^{2}} = -\left[4\frac{R-b}{R} + \frac{4\omega_0^2}{\Omega^{2}} \cos 2X\right] \theta.
\end{equation}
The resulting Mathieu parameters for this system can then be identified as
\begin{equation}\label{eq:SmangMathieuParams}
   a=4 (1 - b/R) \mbox{ and  } q=-2\left(\omega_0 /\Omega\right)^2.
\end{equation}
The interesting aspect of this system is that only the $q$-parameter depends on the rotational frequency of the hoop. The $a$-parameter, on the other hand, varies only with the ratio $b/R$ characterizing the position of the horizontal axis of rotation relative to the hoop center.  For a given value of $b/R$ the parameter $a$ will be constant. Then, by varying the rotational frequency $\Omega$, one can adjust the parameter $q$ from $q \rightarrow 0$ as $\Omega \rightarrow \infty$ to $q \rightarrow \infty,$ as $\Omega \rightarrow 0$.  It is clear from Fig.~(\ref{fig:mathieuPlot}) that the choice of the axis of rotation for the hoop will have a significant impact on the stability of the system.

For example, the choice of $b = 0$ means that the hoop rotates about its diameter. In this case $a = 4$. This line is plotted on the Mathieu stability plot in Fig. (\ref{fig:mathieuPlot}). The remarkable feature of the system in this case is that it is unstable for all possible rotational frequencies, in spite of the fact that the centrifugal force points towards equilibrium and becomes arbitrarily large as the rotational frequency is increased. This highlights the fundamental importance of the time-dependent portion of the Mathieu equation in determining the stability of the solution.

A second example of interest is for the case when $b = R$ so that the axis of rotation passes through one of the equilibrium points on the hoop, the centrifugal force on the bead is minimized, and $a = 0$. This line in $a$ - $q$ space is also shown in Fig.~(\ref{fig:mathieuPlot}). In this case, stability is achievable for sufficiently high rotational rates, $\Omega$, such that $|q|<0.908$. This corresponds to a minimum drive frequency of \begin{equation}\label{eq:minomega}
\Omega_{min} =  1.48 \omega_0.
\end{equation}
In the small angle approximation made above, the system remains stable for arbitrarily high rotational frequencies above the required minimum. As will be shown in Sec.~(\ref{sec:rfPaul}), $b = R$ is the relevant hoop and bead configuration when comparing the behavior of this trap to rf Paul ion traps. The validity of the equation for this configuration is examined in the limit of large $\Omega$ in Sec.~(\ref{sec:smallanglebR}).  As can be seen from Fig.~(\ref{fig:mathieuPlot}), the motion of the bead on the hoop will be stable over the greatest range of rotational frequencies for $a \approx 3$, or when $b \approx 1/4 R.$

\subsection{\label{sec:pseudo} Pseudo-potential approximation}

An important simplification can be made in the analysis of motion described by the Mathieu equation when the parameters $a$ and $q$ are both small. For example this would apply to the hoop trap described in the previous sections  for the stable regime when $b=R,$ so that $a=0$ and $|q| \ll 1$. In particular, $|q| \ll 1$ implies that $\Omega \gg \omega_0$ (see Eqs.~\ref{eq:natfreq} and \ref{eq:SmangMathieuParams}). As a result, the two natural time scales in the problem become so different that the motions associated with them can be decoupled. By averaging over the fast time scale, the ponderomotive force acting on the bead is all that remains. Since the ponderomotive force in the Mathieu equation is linear in the particle displacement, a harmonic pseudo-potential results.  In this pseudo-potential, the particle undergoes a relatively slow oscillation at the frequency $\omega_s$, known as the secular frequency. The faster, small amplitude jitter known as the micromotion, having a frequency tied to the drive frequency $\Omega$, is effectively superimposed on this motion.\cite{ref:dehmelt} The pseudo-potential approximation can be obtained analytically from the general solution of the Mathieu equation in the limit $a,|q|\ll 1$.

The general solution for the motion of a particle described by the Mathieu equation follows from the Floquet theorem,\cite{ref:McLachlan2}
\begin{eqnarray}\label{eq:gensoln}
\theta(X) &=& A e^{i\beta X}\sum_{n = -\infty}^\infty C_{2n}e^{i2nX} \\
&+& B e^{-i\beta X}\sum_{n = -\infty}^\infty C_{2n}e^{-i2nX},\nonumber
\end{eqnarray}
where $A,B$ are arbitrary constants, while $\beta$ and $C_{2n}$ are functions of $a$ and $q$ only. Since the terms inside the sums are periodic functions, the character of the solution is determined by the parameter $\beta$. Complex or imaginary values of $\beta$ give rise to the unstable solutions of the Mathieu equation. Real values of $\beta$ give rise to the stable solutions. Integral values of $\beta$ correspond to the $\pi$- or $2\pi$-periodic solutions of the Mathieu equation. As discussed in Sec.~(\ref{sec:smallanglegen}), these periodic solutions mark the boundary between the stable and unstable regimes in $a$ - $q$ parameter space.

When $\beta$ is real, it is possible to obtain two independent solutions of the Mathieu equation such that
\begin{equation}\label{eq:gensolncese}
\theta(X) = A ce_\beta(q,X) + B se_\beta (q,X),
\end{equation}
where $A$ and $B$ are again real constants, and $ce_\beta$ and $se_\beta$ are power series in the parameter $q$. These solutions reduce to the periodic solutions $\cos \sqrt{a} X$ and $\sin \sqrt{a} X$, respectively, in the limit $q \rightarrow 0.$ They are normalized by convention so that the coefficient of the terms $\cos \beta X$ or $\sin \beta X$ is unity always. In general,\cite{ref:McLachlan3}
\begin{eqnarray}
ce_\beta (X,q) &=& \cos \beta X + \sum_{r=1}^\infty q^r c_r(X),\label{eq:gensolnreal1} \\
se_\beta (X,q) &=& \sin \beta X + \sum_{r=1}^\infty q^r s_r(X), \label{eq:gensolnreal2}\\
a &=& \beta^2 + \sum_{r=1}^\infty \alpha_r q^r. \label{eq:gensolnreal3}
\end{eqnarray}
Since the pseudo-potential approximation applies in the case for $|q| \ll 1,$ approximate solutions to the lowest available order in $q$ are sought.
\begin{eqnarray}
ce_\beta (X,q) &\approx& \cos \beta X + -\frac{q}{4}\left[\frac{\cos (\beta+2)X}{(\beta +1)} - \frac{\cos (\beta-2)X}{(\beta -1)}\right],\label{eq:pseudosoln1} \\
se_\beta (X,q) &\approx& \sin \beta X + -\frac{q}{4}\left[\frac{\sin (\beta+2)X}{(\beta +1)} - \frac{\sin (\beta-2)X}{(\beta -1)}\right], \label{eq:pseudosoln2}\\
a  &\approx& \beta^2 + \frac{q^2}{2(\beta^2-1)}. \label{eq:pseudosoln3}
\end{eqnarray}
For $q$ small, the dominant solution has the dimensionless frequency $\beta$. Much smaller oscillations of relative amplitude $q/4(\beta+1)$ and $q/4(\beta -1)$ are superimposed on the dominant solution at the sideband frequencies of $\beta \pm 2$, respectively. The dimensionless frequency $\beta$ is found from Eq.~\ref{eq:pseudosoln3}. In the specific case when $b = R$ so that $a = 0$, we have:
\begin{equation}\label{eq:secularfreq1}
\beta \approx q/\sqrt{2}.
\end{equation}

In this limit, we obtain for the case of the bead starting from rest at $\theta = \theta_i$ the expression for the displacement of the particle in the pseudo-potential approximation \cite{ref:Leibfried}
\begin{equation} \label{eq:pseudosoln}
\theta(X) \approx \theta_i \cos (q/\sqrt{2} X) \left(1 - \frac{q}{2}\cos(2X)\right).
\end{equation}
As expected, we have a dominant solution at the secular frequency, and a smaller amplitude, higher frequency motion superimposed on it. Recalling that $X = \Omega t/2$ and $|q| = 2(\omega_0/\Omega)^2,$ the secular frequency for the trapped particle in this approximation is then found to be,
\begin{equation}\label{eq:secfreq}
\omega_s \approx \frac{\omega_0^2}{\sqrt{2} \Omega}.
\end{equation}
Note that the amplitude of the micromotion also depends inversely on the drive frequency $\Omega$ through the parameter $q$. The greater $\Omega$ is than $ \omega_0,$ the slower and more dominant the secular motion becomes, so that
\begin{equation}\label{eq:pseudoharmonic}
\theta(X) \approx \theta_i \cos (\omega_s t)
\end{equation} In the pseudo-potential approximation, therefore, we have the following expressions for the time-averaged ponderomotive force, $F_p$ and the pseudo-potential, $\psi$:
\begin{eqnarray}\label{eq:pseudoforce}
F_p &=& -m \omega_s^2 \theta \\
\label{eq:pseudopotential}\psi &=& \frac{1}{2} m \omega_s^2 \theta^2.
\end{eqnarray}

\subsection{\label{sec:rfPaul} Relation to the rf Paul ion trap}

\begin{figure}
 \includegraphics[width=2.5in]{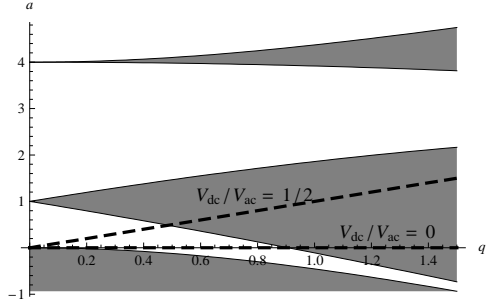}\\
 \caption{Stability plot of the Mathieu parameters $a$ vs. $q$ as it relates to an rf Paul trap. The portion of this parameter space explored by an ion in an rf trap depends on the ratio of the DC to AC voltage, and $\Omega$, the angular frequency of the AC field. For a fixed value of $V_{dc}/V_{ac},$ the possible values for the parameters fall on a line of slope $2 V_{dc}/V_{ac}$ in $a$ - $q$ space. As the frequency $\Omega$ is increased, the value of $q$ decreases until the limit of $q=0$ is reached for an infinitely large $\Omega.$ The lines associated with two possible configurations of the trap are indicated on the plot. The case when $V_{dc} = 0$ corresponds to an ion trapped in a hyperbolic rf Paul trap.}\label{fig:mathieuPlot2}
\end{figure}

The equation of motion along the {\em z}-axis of a particle in an rf Paul hyperbolic trap is\cite{ref:Ruby}
\begin{equation}\label{eq:Rftrap}
    \frac{dz^{2}}{dt^{2}} = -\frac{e}{m}[V_{dc} - V_{ac} \cos \Omega t]\frac{z}{z_{0}^{2}},
\end{equation} where $V_{dc}$ is the constant electric potential in the Paul trap, $V_{ac}$ is the sinusoidally-varying electric potential, and $z_0$ is the distance the cap electrode is away from the trap center along the $z$ axis.
Taking this equation and substituting in $2X=\Omega t$, it can be rewritten in the canonical form of the Mathieu equation
\begin{equation}\label{eq:RftrapMathieu}
    \frac{d^{2}z}{dX^{2}} = -\left[\frac{4 e V_{dc}}{\Omega^{2} m z_{0}^{2}} - \frac{4 e V_{ac}}{\Omega^{2} m z_{0}^{2}} \cos 2X\right] z
\end{equation}
with parameters
\begin{equation}\label{eq:rftrapMathieuParams}
   a=\frac{4 e V_{dc}}{\Omega^{2} m z_{0}^{2}} \mbox{ and  } q=\frac{2 e V_{ac}}{\Omega^{2} m z_{0}^{2}}.
\end{equation}
For a given ion, trap size and AC and DC voltages applied to the electrodes, the motion of the ion will satisfy the Mathieu equation, and the Mathieu parameters will again fall along a line in $a$ - $q$ space. This line is not horizontal, but has a fixed slope equal to $2V_{dc}/V_{ac}$ and a fixed $a$-intercept at $a = 0$. This intercept is reached as $\Omega \rightarrow \infty.$ The allowed parameter space for two possible ion trap configurations as determined by the ratio of DC to AC voltage is shown in Fig.~(\ref{fig:mathieuPlot2}).

The motions of the trapped bead and of the trapped ion can be compared in terms of their respective $a$ and $q$ parameters. Standard operating procedures for hyperbolic rf Paul traps set $V_{dc}=0$, making $a=0$ in Eq.~(\ref{eq:rftrapMathieuParams}). See also Fig.~(\ref{fig:mathieuPlot2}). For the equivalent condition in the hoop trap, one simply sets $b=R$ (see Eq.~(\ref{eq:SmangMathieuParams})). What is more, in both cases the $q$-parameter is a function of the inverse-square of the driving frequency, $\Omega.$ Therefore, for both traps the equations of motion are found to lie along the  line $a=0$ in $a$ - $q$ parameter space, with stable, trapped motion found for values of $q<0.908,$ corresponding to drive frequencies $\Omega$ above some minimum value.

In this sense, the mechanical ponderomotive trap developed here can serve as an excellent illustration of the rf trapping mechanism used in ion traps. The motion of the trapped bead in the small angle approximation explores exactly the same parameter space of the Mathieu equation. The stability of the motion depends on the driving frequency in the same way. And a direct correlation can be drawn between the time-varying electric field in the hyperbolic rf Paul trap and the time-varying gravitational field the bead sees on the rotating hoop. Setting the $q$-parameters for the two cases equal, a comparative relationship between the role played by the two fields is obtained
\begin{equation}\label{eq:Mathieuparamscomp}
    \frac{ eV_{ac}/(mz_0)}{z_{0}} = \frac{g}{R} = \omega_0^2,
\end{equation}
in terms of a ratio of the acceleration experienced by the particle from each field to the relevant distance characterizing the trap size. This corresponds to the natural frequency of each trap, against which the drive frequency must be compared. The equation of the motion of the trapped particle in both traps can therefore be written in the following form:
\begin{equation}\label{eq:Smangbr}
    \ddot{\theta} \approx -\left[\omega_0^2 \cos (\Omega t)\right] \theta.
\end{equation}

\begin{figure}
 \includegraphics[width=2.5in]{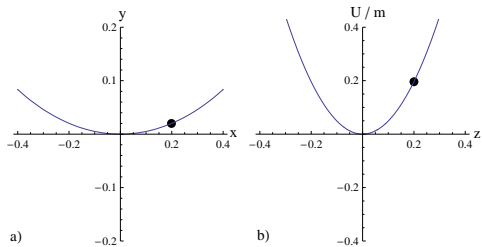}
 \caption{Comparison of the bead on a rotating hoop to the rf Paul trap. Fig. a) displays the instantaneous position of a bead on a rotating hoop of $R = 1.0$  m. Fig. b) displays the instantaneous potential per mass, $U/m$, along one axis of an rf Paul trap and how the ion rides along the surface of that potential.}\label{fig:potentialplot}
\end{figure}

What is more, the spatial and temporal variation of the gravitational field component for small displacements from equilibrium, $g \theta \cos (\Omega t)$,  exactly corresponds to the component of the AC field that the ion sees in the rf trap, $E_z z \cos (\Omega t)$. This means that the instantaneous potential energy for the two traps near the trap center has the same, quadratic form. Since the potential energy in the hoop trap (ignoring the contribution of the centrifugal force) is given by the vertical distance from the bead to the axis of rotation, the rotating hoop displays a real-time plot of the potential energy governing the bead's motion as a function of the horizontal distance. The bead's location along the hoop is therefore a real-time plot of the bead's instantaneous potential energy vs. its instantaneous horizontal position. The superposition of these two `plots' shows the bead moving along the instantaneous potential energy surface within the trap. It is in precisely this manner that the motion of an ion in an rf Paul trap is most easily understood (See Fig.~\ref{fig:potentialplot}).
Therefore, this mechanical ponderomotive trap not only reproduces the correct particle dynamics seen in an rf Paul ion trap, but also serves to illustrate vividly the temporal and spatial variation of the potential associated with the field that creates the rf Paul trap.

\subsection{\label{sec:smallanglebR} The small angle limit for $b = R$ and $\Omega$ large}

When $b=R$, the small angle approximation used above is no longer valid for very large values of $\Omega$, since it ignores entirely the contribution from the centrifugal force. As the rotational rate $\Omega$ becomes very large, the centrifugal force contribution seen in Eq.~(\ref{eq:Fullsln}) becomes a significant factor  even at small displacements, and thus we are compelled to expand the small angle approximations to the next order. Keeping those terms in Eq.~(\ref{eq:Fullsln}) dependent on $\theta$ to the 3rd order and multiplied by $\Omega^2$ gives
\begin{equation}\label{eq:SmangExpand2}
    \ddot{\theta} \approx -\left[\omega_0^2 \cos(\Omega t) \right]\theta + \left[\Omega^{2}/2\right] \theta^{3}.
\end{equation} Equation~(\ref{eq:SmangExpand2}) contains, in addition to the time-dependent ponderomotive term, a third-order term in $\theta$ arising from the time-independent centrifugal force on the bead. The positive sign means that, for $b = R$, the centrifugal force provides a continual push on the bead away from the equilibrium point. So, when the rotational frequency $\Omega$ becomes sufficiently large, the centrifugal force can have a significant impact relative to the gravitational force, changing the motion of the object from stable to unstable.

In the pseudopotential approximation this can be understood as the emergence of an inverted quartic potential term such that, $U'$, the overall time-averaged potential governing the particle's secular motion becomes anharmonic and turns over at a value of $\theta$ that is inversely related to the angular velocity, $\Omega$,
\begin{equation}\label{eq:quarticterm}
U' = \frac{1}{2}m\omega_s^2 \theta^2 - \frac{1}{8}m\Omega^2 \theta^4.
\end{equation}
The potential $U'$ turns over when its slope is zero, such that
\begin{equation}\label{eq:turnover}
m\omega_s^2 \theta - \frac{1}{2}m\Omega^2 \theta^3 = 0.
\end{equation}
Substituting in the expression for the secular frequency from Eq.~(\ref{eq:secfreq}), one can solve for the maximum value of $\theta_m$ for which the time-averaged potential is still trapping,
\begin{equation}\label{eq:thetam}
\theta_m = \omega_0^2/ \Omega^2.
\end{equation}
The effect of the centrifugal force, therefore, is to impose an upper limit on the angular frequency of the hoop as a function of the initial displacement of the bead, $\theta_i$, for which our device will trap the particle,
\begin{equation}\label{eq:Omegamax}
(\Omega/\omega_0)_{\mbox{max}} = 1/\sqrt{\theta_i},
\end{equation}
in addition to the lower frequency limit already obtained in Eq.~(\ref{eq:minomega}) from the stability characteristics of the Mathieu equation.

These two limits for stable motion can be seen in
\begin{figure}
 \includegraphics[width=2.5in]{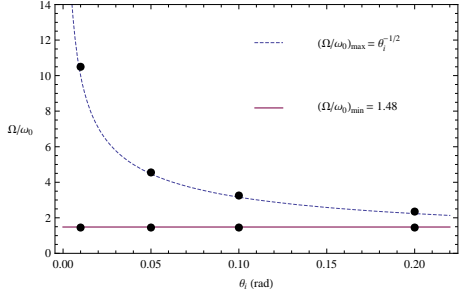}
  \caption{Plots of maximum and minimum stability values of $\Omega/\omega_0$ vs. $\theta_i$, the initial displacement of the bead. The plots show the theoretical maximum and minimum stability limits for the bead (dashed and solid lines), as well as the observed stability limits from the full numerical solution (points). The agreement between the predictions of the lowest order approximations and the full solution is very good. }\label{fig:fullslnlimits}
\end{figure}
Fig.~(\ref{fig:fullslnlimits}), where the theoretical minimum and maximum stability limits for a bead starting from rest are compared against those obtained numerically from the full equation of motion, Eq.~(\ref{eq:Fullsln}). Over the range of initial displacements tested in the numerical analysis, there was no difference observed in the onset of stability for the third-order equation and the full equation of the bead's motion. The agreement between the theoretically-obtained approximate limits and those from the numerical solution of the full equation is remarkably good, and validates the usefulness of the pseudo-potential approximation in analyzing the bead's motion. As a result, it is possible to relate the behavior of this system to a large number of disparate physical systems whose behavior can be described in terms of an anharmonic potential such as found in Eq.~(\ref{eq:quarticterm}).\cite{ref:Rau}

\subsection{\label{sec:inverpend} Relation to the inverted pendulum}

The behavior and motion of the inverted pendulum has also been compared to the rf ion trap. The equation of motion for the inverted pendulum is\cite{ref:vpend}
\begin{equation}\label{eq:invpend}
    \ddot{\theta} =
     (3/2l) \left[g - b \Omega^{2} \cos \Omega t\right] \theta
\end{equation}
for a pendulum of length $l$ and amplitude of displacement of the pivot point, $b$. Again, the canonical form of the Mathieu equation can be obtained by substituting $X=\Omega t/2$ into Eq. (\ref{eq:invpend}). Thus,
\begin{equation}\label{eq:invpendmathieu}
    \frac{d^{2}\theta}{dX^{2}} = -\left[-(6g/(l\Omega^{2})) + (3b/l) \cos 2X\right] \theta
\end{equation}
with parameters
\begin{equation}\label{eq:invpendMathieuParams}
   a=- 6g/(l\Omega^{2}) \mbox{ and  } q=-3b/l.
\end{equation}

\begin{figure}
 \includegraphics[width=2.5in]{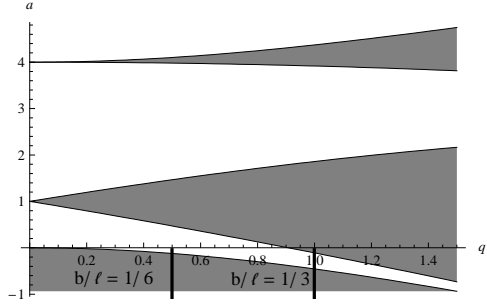}
 \caption{Stability plot of the Mathieu parameters $a$ vs. $q$ as it relates to the inverted pendulum. The portion of this parameter space explored by the vertical pendulum depends on the ratio $b/\ell$, the displacement of the connection point to the length of the pendulum, and $\Omega$, the angular frequency at which the pendulum is driven. For a fixed value of $b/\ell,$ the value of $q$ is fixed. The possible values for the parameters fall on a vertical line below the $q$-axis in $a$ - $q$ space. As the frequency $\Omega$ is increased, the value of $a$ approaches 0. The lines associated with two possible configurations of the pendulum are indicated on the plot. }\label{fig:mathieuPlot3}
\end{figure}

As with the hoop trap and the rf trap, the geometry of the vertically-driven pendulum constrains the possible values of the Mathieu parameters to a line in $a$ - $q$ space (See Fig.~\ref{fig:mathieuPlot3}).
However, the behavior of the inverted pendulum is different from  that of either the hoop trap or the rf trap. First of all, the geometry constrains the $q$-parameter to be constant, while the $a$-parameter varies with $\Omega.$ Therefore, the lines are vertical in $a$ - $q$ space, which is never the case in an rf trap.  Secondly, the $a$-parameter is negative, greatly restricting the possibility for stable motion to occur. It is possible to choose the value of $q$ so that the stability of the pendulum as a function of frequency mimics the stability of ions in rf traps. Fundamentally, however, the visual comparison of the ponderomotive potential controlling the inverted pendulum's behavior to the potential seen by the ion in the rf Paul trap is difficult to make.

It should be pointed out that the behavior described in Sec.~\ref{sec:theory} could also be interpreted as belonging to a variation of the inverted pendulum described above, where the bead is the pendulum bob and the circular path represented by the hoop is fixed by a rigid, massless rod of length $R$. In this case, rather than driving the pivot point of the rod in the vertical direction, the point is made to rotate about a horizontal axis a distance $b$ below it. As viewed in the lab frame, stable motion would involve the connection point oscillating up and down and the pendulum bob alternating between being located above and below the pivot point. This model was considered as an alternative for a demonstration model of the proposed system.  However, it was ultimately rejected because of the concern  about excessive torque and friction at the necessary size and speed of rotation of the system.

\subsection{\label{sec:noisefriction} Effects of Noise and Friction}

As a preliminary to the construction of an apparatus that displays the proposed hoop trap behavior, numerical simulations were carried out to determine the effects of noise and friction on the motion of the bead in the hoop trap.

Because ponderomotive traps depend on both the temporal and spatial variation of the trapping field, fluctuations in both aspects of the field must be considered. For the proposed hoop trap, spatial variations in the trapping field could arise from fluctuations in the shape of the hoop due to twisting or bending, while temporal variations could arise from fluctuations in the rotational frequency of the motor. It was considered that the latter would be much more difficult to control than the former, so the numerical studies carried out focused on a ``pink noise'' variation in the rotational frequency, $\Omega,$ such that $\Omega(t) = \Omega_0 ( 1+ \varepsilon (t)).$ The ``pink noise'' was introduced by generating a discrete noise time sequence that displayed a 1/f dependence on the frequency in its spectral energy density over a five-decade range of frequencies around the drive frequency of the system.\cite{ref:noise} The sequence was approximated as continuous for shorter time scales. It was found that the presence of such noise was destabilizing for frequency variations on the order of 1\% of the drive frequency or greater. As a result, it was determined that investment in a high performance, high rpm motor would be needed for carrying out the demonstration successfully.

Frictional damping is a crucial feature in any kind of trap. The concern with the hoop trap was that the friction between the bead and hoop would be so great as to prevent entirely observation of any frequency-dependent effects on the motion. To account for the effects of friction, Eq.~(\ref{eq:Fullsln}) can be modified with a term proportional to the normal force of the hoop on the bead. The normal force of the hoop arises in reaction to the gravitational and centrifugal forces acting on the bead in the rotating frame of the hoop.  The equation of motion for the bead becomes
\begin{eqnarray}\label{eq:Fullslnfriction}
\ddot{\theta} &=& -[\Omega^{2} (\cos \theta - 1) + \omega_0^2 \cos \Omega t] \sin \theta - \mu \mbox{Sign}[\dot{\theta}] *\\
&*&  \sqrt{ \left\{ \Omega^2 (\cos \theta - 1) + \omega_0^2 \cos \Omega t  \right\}^2 \cos^2 \theta+ \omega_0^4 \sin^2 \Omega t } .\nonumber
\end{eqnarray}
This equation is appropriate for both sliding and rolling friction. Numerical simulations showed that the motion of the bead was always stable for $\mu > 0.01$. A sliding friction coefficient of $\mu < 0.01$ is not achievable for ordinary materials. Therefore, based on the numerical results, the decision was made to modify the design and have the bead roll rather than slide along the hoop.

\subsection{\label{sec:rolling} Rolling instead of sliding}

A bead that rolls along the hoop instead of slides will have an identical potential energy term but a slightly modified kinetic energy term in the Lagrangian function. The kinetic energy is now described as
\begin{equation}\label{eq:kinenRoll}
    T = {\scriptstyle \frac{1}{2}}  m (R \cos \theta - b)^{2} \Omega^{2} + {\scriptstyle \frac{7}{10}}  m R^{2} \dot{\theta}^{2}
\end{equation}
Using the potential from Eq. (\ref{eq:poten}), the Lagrangian becomes
\begin{eqnarray}\label{eq:LagrRoll}
    {\cal L} &=& {\scriptstyle \frac{1}{2}} m (R \cos \theta - b)^{2} \Omega^{2} + {\scriptstyle \frac{7}{10}} m R^{2} \dot{\theta}^{2} \\
&+& m g (R \cos \theta - b) \cos \Omega t.
\end{eqnarray}
Applying $d{\cal L}/d\theta = (d/dt) (d{\cal L}/d\dot{\theta})$ to Eq. (\ref{eq:LagrRoll}) results in the equation of motion for a bead rolling along a rotating hoop
\begin{equation}\label{eq:FullslnRoll}
    \ddot{\theta} = -{\scriptstyle \frac{5}{7}}[\Omega^{2} [(R \cos \theta - b)/R] + \omega_0^2 \cos \Omega t] \sin \theta.
\end{equation}
Allowing the bead to roll rather than slide along the hoop results in the presence of an overall factor of $5/7$ in the expression for the acceleration of the center of mass of the bead.  In the small angle approximation, this results in a modification of the Mathieu parameters, so that they become
\begin{equation}\label{eq:SmangMathieuParamsRoll}
   a={\scriptstyle \frac{20}{7}} (1 - b/R) \mbox{ and  } q=-{\scriptstyle \frac{10}{7}}\frac{\omega_0^2}{\Omega^{2}}.
\end{equation}
In the special case that $b= R$, one could simply redefine the natural frequency as
\begin{equation}
\omega_0' = \sqrt{5g/7R},
\end{equation}
which is the natural frequency for a ball rolling along the bottom of a stationary circular hoop. In this sense, the analysis done in Sec.~(\ref{sec:smallanglebR}) remains valid, as long as one substitutes in the modified natural frequency, $\omega_0'$. However, since the natural frequency of the system is reduced for the same radius of hoop, the minimum frequency of rotation necessary for stability is also reduced by a factor of $\sqrt{5/7}$, so that
\begin{equation}\label{eq:minomegaroll}
    \Omega_{min} = 1.14 \omega_0.
\end{equation}
The secular frequency for a rolling ball in the pseudo-potential approximation, which depends on the square of the natural frequency is therefore reduced by a factor of $5/7$,
\begin{equation}\label{eq:seqfreqroll}
\omega_{s} = {\scriptstyle \frac{5}{7}} \left( \frac{\omega_0^2}{\sqrt{2}\Omega}\right).
\end{equation}

In general, the rolling ball in the rotating hoop will display all of the features of the 1-D ponderomotive traps discussed above. With a sufficiently smooth bead and rolling surface, it is possible to reduce the effects of friction enough that the transition between stable and unstable motion becomes clearly visible, and the secular frequency predicted here is in fact observed, as demonstrated in the experimental results section below.

\section{\label{sec:demon} DEMONSTRATION APPARATUS}

\subsection{\label{sec:design} Design}

\begin{figure}
 \includegraphics[width=2.5in]{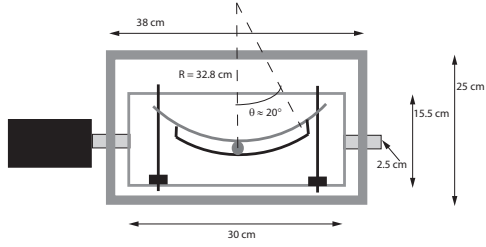}\\
 \caption{Schematic of the hoop trap demonstration apparatus. All dimensions are given in cm. The thickness of the pieces making up the outer box was 1.3 cm, while the thickness of the pieces making up the inner box was 0.6 cm. The width of the pieces for the outer box was 7.5 cm, and the width of the inner pieces was 3.5 cm. }\label{fig:diagram}
 \end{figure}

As was discussed in Sec.~(\ref{sec:noisefriction}), concerns about noise and friction were paramount in developing a demonstration apparatus that would display the features described in this paper. The design for the demonstration apparatus is shown in Fig.~(\ref{fig:diagram}).
 A spherical steel ball-bearing of diameter $d = 0.63$ inches was used as the trapped particle. The bearing was allowed to roll inside of clear, flexible tygon tubing 1.0 inch OD (0.75 in ID) that was tied to a portion of a 26 inch bicycle rim to keep the circular shape and capped at both ends by a rubber stopper. Only a 26 cm portion of the rim was used, which allowed for about a 30$^\circ$-range of motion for the bearing. The choice of the bicycle rim and tygon tubing fixed the radius of the rotating circular hoop at $R = 0.328 m$. This then fixed the value of the natural frequency of the ponderomotive trap
\begin{equation}\label{eq:omega0}
\omega_0 = \sqrt{g/R}  =  5.47 rad/s.
\end{equation}
This sets the requirement for a minimum drive motor frequency of
\begin{equation}\label{eq:minomeganeed}
\Omega_{min} = 1.14\omega_0 = 6.24 rad/s,
\end{equation} for trapping the rolling ball. It was assumed that the presence of a significant amount of friction would make calculating the maximum drive frequency unnecessary.

A rectangular aluminum frame was constructed for the rim from 0.6 cm-thick aluminum plates cut into about 3.5 cm-thick strips. The frame for the rim and tube was 15.5 cm high and 30 cm long. The rim was attached to the frame using two threaded rods that were threaded through the top and bottom of the frame and passed through spoke holes at the ends of the rim section. The rim could be adjusted up and down along the threaded rod using nuts and washers. Shaped pieces of rubber stopper were used to minimize vibrations of the rim. Counter-weights to keep the whole piece balanced about the axis of rotation were attached to the threaded rod on the side of the frame opposite the rim. Two pieces of round aluminum stock 2.5 cm in diameter and 7.5 cm long were then attached to the two sides of the frame to form the axle of the rotating frame for the apparatus. They were to pass through two pump bearings so that the frame could rotate freely. One of the two pieces of round stock was hollowed out on one end to make a sleeve so that it fit over the drive shaft of the motor.

\begin{figure}
 \includegraphics[width=2.5in]{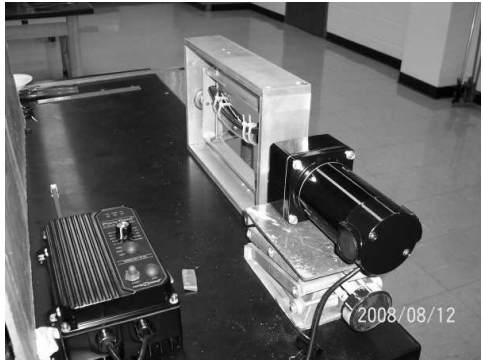}\\
 \caption{Photo of the hoop trap, motor and controller.}\label{fig:photo}
 \end{figure}

The two pump bearings were press-fitted into circular holes in the sides of a stationary outer frame made of 1.3 cm-thick stock cut into strips 7.5 cm wide. The outer frame was 25 cm high and 38 cm long. It was clamped to a table and leveled using shims. The motor was raised to the level of the axle using a lab jack. The motor used was a Bodine 5160 DC inline 1/8 hp motor, with a KB-240 controller. This motor is designed to reach
\begin{equation}\label{eq:maxmotorfreq}
\Omega_{max} = 266 rpm = 27.9 rad/s,
\end{equation} which is well above the minimum frequency predicted in Eq.~(\ref{eq:minomeganeed}). For the experiments described below, the controller's potentiometers were set so as to be able to vary the motor speed from 0 to 21.4 rad/s. Both of these devices were purchased with the help of an undergraduate research grant from the College of Arts and Sciences at WIU. A photo of the working apparatus is shown in Fig.~(\ref{fig:photo}).

\subsection{\label{sec:perform} Performance}

Once the device was constructed, the only significant issue that arose during its operation was the need for careful balancing and leveling of the device. Without proper balancing, the motor and frame tended to shake, especially at higher frequencies. Pieces of lead were attached to the bottom of the frame opposite the rim until the inner frame was well-balanced about the rotational axis. Without proper leveling, the equilibrium point for the rotating system was shifted to the side from the static equilibrium point, and the system became unstable for moderately large rotational frequencies. In effect, the small vertical component of the rotational axis resulted in a component of the centrifugal `field' that pushed the ball away from the bottom of the hoop, acting like an offset DC field in the rf trap (See the appendix). Although the tygon tubing was not perfectly cylindrical due to the two edges of the bicycle rim pressing down on it, the ball bearing was able to roll smoothly in place as the tube rotated around it, and it was able to move along the tube without any significant hindrance.

Once the system was balanced and leveled, the device clearly displayed the characteristic features of a ponderomotive trap. The minimum frequency at which stability could be maintained was investigated by several methods. First, the motor was started from rest with the hoop upright so that the ball was at rest at the bottom of the hoop. The motor was turned on suddenly (adiabatically) and the ball's motion was examined for stability. It took approximately half of one second for the motor to reach full speed. If the motor was turned up to too slow of a rotational speed, the ball quickly became unstable and began rolling from end to end in the closed tygon tube. But if the motor was turned up to a sufficiently high rate of rotation, the ball was trapped at the bottom of the rotating hoop. The lowest frequency at which the ball remained in stable motion was recorded. The second method was to turn the motor on to a higher rotation speed, trap the ball, and then gradually lower the motor speed until the ball became unstable. Both methods gave essentially the same value of
\begin{equation}\label{eq:minomegaexpt}
\Omega_{min} \approx 6.3 rad/s = 1.15 \omega_0.
\end{equation} This value compares remarkably well with the predicted value of $\Omega_{min} = 1.14 \omega_0.$ The ball remains stably trapped up to the highest possible rotational frequency. No data has yet been taken to determine stability as a function of initial displacement. Nevertheless, the apparatus provides a clear demonstration how a ponderomotive trap works, and simple measurements made of the rotational frequency at the onset of stability match the theoretical prediction.

\begin{figure}
 \includegraphics[width=2.5in]{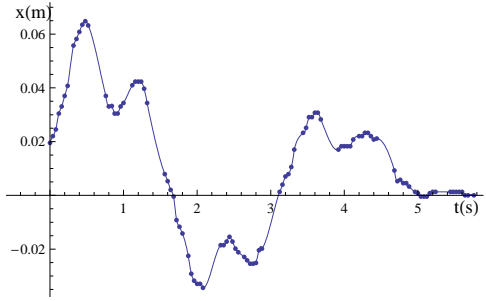}
 \caption{Horizontal position of the center of mass of the ball as a function of time. The rotational frequency of the trap was $\Omega = 8.06 \frac{rad}{s}.$ The data points were obtained using Vernier's Logger Pro 3.6 motion analysis feature. Not all the frames could be used in the analysis because the ball was occasionally obscured by the frame and rim. An interpolating function was created from the data using Mathematica in order to do a spectral analysis of the motion, and is plotted as a solid line. The secular motion and superimposed micromotion are clearly seen. The effects of frictional damping are also visible.}\label{fig:motionplot}
\end{figure}

A second test of the apparatus was whether the trapped ball could display clearly both secular motion and micromotion, and if the measured secular frequency would match the theoretical prediction obtained above. The current design does not provide a simple way of holding the ball at an initial position other than at equilibrium. However, if the hoop is rotating slowly enough, it is possible to perturb the ball using a strong permanent magnet attached to a longer bar magnet and then watch its return to the equilibrium position. This motion was videotaped using a video camera at relatively high shutter speed (1/250th of a second) but standard frame speed (25 fps), and the motion of the ball analyzed frame by frame using the video analysis package in Vernier's LoggerPro 3.6 software. The motor's rotational frequency for this demonstration was also measured using video analysis, and found to be $\Omega = 1.28 rps = 8.06 \frac{rad}{s}.$ Figure~(\ref{fig:motionplot})
shows the horizontal position of the center of mass of the ball as a function of time. Not every frame provided a data point because the presence of the frame and the bicycle rim often obscured the ball's location. Nevertheless, the plot clearly indicates the presence of a slower, secular motion, with a smaller amplitude jitter superimposed on it. This motion was clearly visible on the videotape and, when motion like it was shown in a live demonstration to visiting high school students, it drew a number of ``wows''.

\begin{figure}
 \includegraphics[width=2.5in]{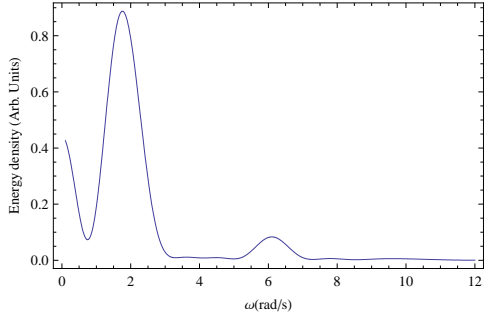}
 \caption{Spectral analysis of the ball's motion. The vertical axis represents the energy in the motion on a scale with arbitrary units. The two major peaks correspond to the secular frequency, $\omega_s = 1.9 \frac{rad}{s},$ and the frequency difference between the rotational frequency and secular frequency of $\omega_{\Delta-} = 6.1 \frac{rad}{s}.$ The frequency sum peak is suppressed due to the relatively large values of $q$ and $\beta$.} \label{fig:spectralanalysis}
\end{figure}

In the pseudopotential approximation, the secular frequency of the motion of the rolling ball in the trap should be given by the formula found in Eq.~(\ref{eq:seqfreqroll})
\begin{equation}\label{eq:secfreqdata}
\omega_s = {\scriptstyle \frac{5}{7}} \frac{\omega_0^2}{\sqrt{2}\Omega} = 1.87 \frac{rad}{s}.
\end{equation}
A spectrum analysis was carried out on the motion using Mathematica 6.0, and is shown in Fig.~(\ref{fig:spectralanalysis}).
The vertical axis indicates the energy density of the signal in arbitrary units. The major features of the spectrum occur at the secular frequency of the motion, with an estimated value of
\begin{equation}\label{eq:secfreqspec}
\omega_s = 1.9 \frac{rad}{s}.
\end{equation} This is in excellent agreement with the predicted value.

The second major feature of the spectrum occurs at $\omega = 6.1 \frac{rad}{s}.$ This peak occurs at the frequency difference between the rotational frequency and the secular frequency, $\omega_{\Delta-} = \Omega - \omega_s = 8.06\frac{rad}{s} - 1.9\frac{rad}{s} = 6.2\frac{rad}{s},$ as expected from the equation of motion in the pseudopotential approximation (See Eq.~\ref{eq:pseudosoln}). Another spectral component at the sum of the frequencies is barely visible at $\approx$ 10 rad/s. This peak is suppressed in the energy spectrum due to the fact that, at the value of $\Omega$ used in this trial, $q = 0.66$, and $\beta \approx 0.46.$ As a result, the term corresponding to the frequency sum in Eq.~(\ref{eq:pseudosoln1}) has an amplitude greatly suppressed relative to that of the frequency difference.  In addition, other factors may have further suppressed this peak, such as friction and deformation of the tubing due to the edges of the rim, or to the plastic ties holding the tube to the rim.

\begin{figure}
 \includegraphics[width=3.0in]{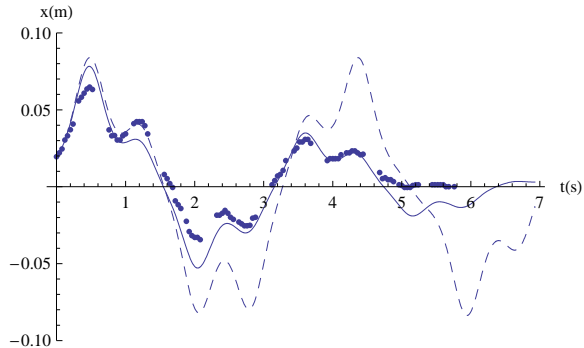}
 \caption{Simulation of the motion of the rolling ball in the hoop trap as compared to the actual recorded motion from Fig.~(\ref{fig:motionplot}). The solid line indicates motion for a system with a rolling friction coefficient of $\mu =0.007.$ The dashed line indicates the motion without friction, as predicted by Eq.~(\ref{eq:FullslnRoll}). The data points were obtained from video analysis of the motion as recorded during the operation of the demonstration apparatus. The time for the plot was offset by $t = 0.304 s$ to account for the initial orientation of the hoop. The initial angle is $\theta_i = 0.060$ rad. The initial angular velocity is $\dot{\theta}_i = 0.19$ rad/s. } \label{fig:simmotion}
\end{figure}

The remarkable agreement between the overall features of the observed motion seen in Fig.~(\ref{fig:motionplot}) and the predicted features from the equation of motion in Eq.~(\ref{eq:FullslnRoll}) extends to the details of the motion of the ball. The data from the video analysis was used to obtain the initial displacement ($\theta_i = 0.06$ rad) and angular velocity ($\dot{\theta} = 0.19$ rad/s) of the ball and also the orientation of the hoop, expressed as an initial time ($t = 0.78\pi/\Omega = 0.304$ s). An estimate for the frictional coefficient ($\mu = 0.007$) was obtained by releasing the ball from a fixed position while the hoop was stationary and upright, observing its motion and then matching that motion with a numerical simulation using Eq.~(\ref{eq:Fullslnfriction}). The resulting prediction of the motion is displayed in Fig.~(\ref{fig:simmotion}). In addition, the motion that would have resulted in the absence of friction is displayed as a dashed line.
 While the motion predicted here agrees well qualitatively with that observed in the actual device, it is apparent that the predicted motion is initially of a larger amplitude than what was observed. Most likely, this is the result of the motion of the ball extending beyond the location of the first plastic tie holding the tube to the bicycle rim. At that point, the tube was deformed slightly inward, causing the ball to slow down both on the way out and on the way back down again. A second factor reducing the apparent amplitude of the motion on the device was the foreshortening of the distances away from the center of the video used for the motion analysis. Nevertheless, the similarities between the two motions are quite impressive, and it is supposed that with minor modifications of the device construction and data acquisition techniques a near perfect match between the predicted and actual motions can be obtained.

\section{\label{sec:conclusions} CONCLUSIONS}

A new design for a one-dimensional ponderomotive trap using a time- and spatially-varying gravitational field has been proposed and constructed. This device consists of a hollow circular hoop with a ball bearing located inside the hoop which is then rotated about a horizontal axis passing through the plane of the hoop. The model suggested can be used to explore experimentally a large range of the parameter space of the Mathieu equation by varying the axis location and the frequency of rotation. In particular, this model serves as a mechanical analog of the rf Paul ion trap in the case when the axis of rotation is tangent to the hoop. A demonstration apparatus was designed and built which displays all the features of an rf Paul trap to a remarkable degree of accuracy. Not only so, but the motion of the ball in the hoop presents a real time `graph' of the instantaneous position of the ball and potential energy curve within the trap that mimics the corresponding graphs in the rf Paul trap.

Not only can this device be used to demonstrate the operation of an rf Paul trap, it can also be used to explore the behavior of solutions to the Mathieu equation over a wide range of the Mathieu parameters, including the surprising case when $b = 0$ for which the motion is always unstable at small angles. Additional modifications to the device will allow for experimental studies on the effect of noise, friction and anharmonic contributions to the potential energy creating the ponderomotive force.
\begin{acknowledgments}
The authors would like to thank Prof. Bill Case for support, encouragement and some very helpful suggestions regarding this work. We would also like to thank Prof. Chris Monroe for one of the author's (JAR) awareness of and interest in rf ion traps. We owe a huge debt of gratitude to Mr. Eldon Hare who helped immeasurably in the construction of the demonstration apparatus. Finally, both authors would like to thank the Undergraduate Research Council of the College of Arts and Sciences at Western Illinois University for their support of this project through an undergraduate research grant.
\end{acknowledgments}

\appendix

\section{\label{sec:appendA}ADVANCED UNDERGRADUATE LEVEL PROBLEMS}

The physical system developed in this paper was originally conceived as a problem for the final exam in the classical mechanics course taught at Western Illinois University. Certain other aspects of this problem have been deliberately left unworked, and can serve as homework problems or test problems in an upper division classical mechanics course.

\subsection{\label{sec:prob1}The pseudo-potential approximation}

Obtain the equation of motion for the bead in the pseudo-potential approximation given in Eq.~(\ref{eq:pseudosoln}), starting from the approximate solutions to the Mathieu equation given in Eqs.~(\ref{eq:pseudosoln1} - \ref{eq:pseudosoln3}).

\subsection{\label{sec:prob2}Motion of the bead on a hoop with a tilted axis of rotation}

This problem analyzes the physical situation mentioned at the beginning of Sec.~(\ref{sec:perform}), in which the apparatus is not properly leveled, so that the axis of rotation is not perfectly horizontal.

\subsubsection{\label{sec:prob2a} The general solution when the axis passes through the center of the hoop}

Consider a bead of mass $m$ free to slide along a hoop of radius $R$. The hoop is rotated at angular velocity $\Omega$ about an axis that passes through the center of the hoop in the plane of the hoop, but which is tilted in the vertical direction by an angle $\alpha$ from the horizontal.

Show that the equation of motion for the bead is then:
\begin{eqnarray}\label{eq:prob2a}
\ddot{\theta} &=& \Omega^2 \cos(\alpha -\theta)\sin(\alpha - \theta) + \\
&+& \omega_0^2(\cos \Omega t \cos \alpha \sin(\alpha -\theta) - \sin \alpha \cos(\alpha-\theta).\nonumber
\end{eqnarray}
Show that for the extreme cases when $\alpha = 0$ and $\alpha = \pi/2,$ the equation reduces to either the problem discussed in this paper ($b = 0$) or to the equation of motion of a bead on a hoop rotating about a vertical axis, as obtained in Taylor.\cite{ref:Taylor}

\subsubsection{\label{sec:prob2b} The limiting case for small angles}

Show that in the limit of $\alpha,\theta \ll 1$,
\begin{equation}\label{eq:prob2b}
\ddot{\theta} = \left(\Omega^2  + \omega_0^2 \cos \Omega t\right) (\alpha -\theta) - \omega_0^2 \alpha .
\end{equation}
Discuss this solution. What happens to the equilibrium point of the gravitational force in this case? What happens to the equilibrium point of the centrifugal force?

\subsubsection{\label{sec:prob2c} The general solution when the axis passes through the bottom of the hoop}

Show that for the case when the axis of rotation is lowered a distance $b = R$ so that it intersects the hoop at its bottom, the equation of motion for the bead becomes:
\begin{eqnarray}\label{eq:prob2c}
\ddot{\theta} &=& \Omega^2( \cos(\alpha -\theta)-\cos\alpha)\sin(\alpha - \theta) + \\
&+& \omega_0^2(\cos \Omega t \cos \alpha \sin(\alpha -\theta) - \sin \alpha \cos(\alpha-\theta).\nonumber
\end{eqnarray}
Use Mathematica or another computer algebra system to find the solution numerically in order to examine the influence of the tilt angle $\alpha$ on the solution. Use the parameters, $\Omega = 10, \omega_0 = 5, \theta_i = 0.1,$ and try $\alpha = 0, 0.01,$ and $0.1$ as a start. Try to find the maximum $\alpha$ for which the bead will remain localized. Explain why, for larger $\alpha$, the bead is inevitably pushed toward the lowered end of the apparatus. What happens as you increase the angular frequency, $\Omega$?


\begin{thebibliography}{00}
\bibitem{cit:massspec}Raymond E. March, ``Quadrupole Ion Trap Mass Spectrometry:
Theory, Simulation, Recent Developments and Applications,'' Rapid Commun. Mass Spectrom. {\bf 12}, 1543–-1554 (1998).
\bibitem{cit:quinfo}J. I. Cirac, and P. Zoller, ``Quantum computation
with cold, trapped ions,'' Phys. Rev. Lett., {\bf 74} (20), p. 4091 (1995).
\bibitem{ref:dehmelt}Hans G. Dehmelt, ``Radiofrequency spectroscopy of stored ions I: Storage,'' Adv. At. Mol. Phys. {\bf 3}, 53--72 (1967).
\bibitem{ref:Paul} Wolfgang Paul, ``Electromagnetic traps for charged and neutral particles,'' Rev. Mod. Phys., {\bf 62} (3), 531--540 (1990).
\bibitem{ref:vpend}M. H. Friedman, J. E. Campana, L. Kelner, E. H. Seeliger and A. L. Yergey, ``The inverted pendulum: A mechanical analog of the quadrupole mass filter,'' Am. J. Phys. {\bf 50} (10), 924--931 (1982).
\bibitem{ref:saddle}W. Rueckner, J. Georgi, D. Goodale, D. Rosenberg and D. Tavilla, ``Rotating saddle Paul trap,'' Am. J. Phys. {\bf 63} (2), 186--187 (1995).
\bibitem{ref:Rau} A. R. P. Rau, ``The asymmetric rotor as a model for localization,'' Rev. Mod. Phys., {\bf 64} (2), 623--632 (1992).
\bibitem{ref:monroe} Chris Monroe, private communication (2004).
\bibitem{ref:lanlif} L. D. Landau and E. M. Lifshitz, \textsl{Mechanics}, (Pergamon Press, NY, NY, 1976) 3rd ed., pp. 11-12.
\bibitem{ref:Taylor} John R. Taylor, \textsl{Classical Mechanics}, (University Science Books, Sausalito, CA, 2005) 1st ed., pp. 260--264.
\bibitem{ref:Ruby} Lawrence Ruby, ``Applications of the Mathieu equation,'' Am. J. Phys., {\bf 64} (1), 39--44 (1996).
\bibitem{ref:McLachlan1}Norman W. McLachlan, \textsl{Theory and Application of Mathieu Functions}, (Dover, NY, NY, 1964) 1st ed., p. 40.
\bibitem{ref:McLachlan2}Ibid., p. 59.
\bibitem{ref:McLachlan3}Ibid., p. 19.
\bibitem{ref:Leibfried} D. Leibfried, R. Blatt, C. Monroe and D. Wineland, ``Quantum dynamics of single trapped ions,'' Rev. Mod. Phys., {\bf 75} (1), 281--324 (2003).
\bibitem{ref:noise} N. Jeremy Kasdin, ``Discrete Simulation of Colored Noise
and Stochastic Processes and $1/f^\alpha$ Power Law Noise Generation,'' Proc. IEEE, {\bf 83} (5), 802--827 (1995).
\end{thebibliography}
\end{document}